\documentclass[journal]{IEEEtran} 

\usepackage{color}
\usepackage[table]{xcolor}
\usepackage{xcolor}
\usepackage{amsmath}
\usepackage{multirow}
\usepackage{amssymb}

\usepackage{algorithm}
\usepackage[noend]{algpseudocode}

 \usepackage{amsthm}

\usepackage{array}

\usepackage{subcaption}
\usepackage{graphicx}

\usepackage{soul}
\usepackage{cite}
\usepackage{adjustbox}
\usepackage{xcolor,cite,etoolbox}

\begin{document}
\title{HAPS for 6G Networks: Potential Use Cases, Open Challenges, and Possible Solutions}
\author{Omid~Abbasi\IEEEauthorrefmark{1}, Animesh Yadav\IEEEauthorrefmark{1}, Halim~Yanikomeroglu, Ngoc Dung Dao, Gamini Senarath, and Peiying Zhu  \thanks{This work was supported by Huawei Canada Co.,
Ltd. \\
\IEEEauthorrefmark{1}Authors have contributed equally to this research.\\
Omid Abbasi, Animesh Yadav, and Halim Yanikomeroglu are with
Carleton University, Canada; Ngoc Dung Dao, Peiying Zhu, and Gamini Senarath are with Huawei Technologies Canada Co., Ltd.}}
\maketitle

\begin{abstract}
High altitude platform station (HAPS), which is deployed in the stratosphere at an altitude of 20-50 kilometres, has attracted much attention in recent years due to their large footprint, line-of-sight links, and fixed position relative to the Earth. Compared with existing network infrastructure, HAPS has a much larger coverage area than terrestrial base stations and is much closer than satellites to the ground users. Besides small-cells and macro-cells, a HAPS can offer one mega-cell, which can complement legacy networks in 6G and beyond wireless systems.
This paper explores potential use cases and discusses relevant open challenges of integrating HAPS into legacy networks, while also suggesting some solutions to these challenges. The cumulative density functions of spectral efficiency of the integrated network and cell-edge users are studied and compared with terrestrial network. The results show the capacity gains achieved by the integrated network are beneficial to cell-edge users.  Furthermore, the advantages of a HAPS for backhauling aerial base stations are demonstrated by the simulation results.
\end{abstract}


%
\IEEEpeerreviewmaketitle

\section{Introduction}

With the commercialization of 5G in the early phases, the attention of academia and industry is shifting to forecasting the shape of the next-generation wireless network standard. Research for sixth generation (6G) wireless network is in full swing \cite{Uusitalo20216GVV,Saad-Bennis-Chen-NM-2020}. 
\textcolor{black}{It is envisioned that 6G will provide an optimum user experience for all through hyper-connectivity (involving humans and everything) and the development of a ubiquitous intelligent mobile society. Several studies (e.g., \cite{Uusitalo20216GVV,Saad-Bennis-Chen-NM-2020}) have predicted possible key performance indicators (KPIs) of 6G systems and provided a summary of enabling technologies needed to realize the KPIs.}

\textcolor{black}{Among other enabling technologies, the} integration of terrestrial networks (TNs) with NTNs for 6G is gaining tremendous attention from industry and academia \cite{Alouini}. \textcolor{black}{NTN primarily consists of two layer of networks, i.e., satellite and aerial networks. The integrated TN and NTN is referred to as a three-layered vertical HetNets (vHetNets) \cite{alzenad2018IEEEcm}.}  The vHetNet structure is envisioned to provide pervasive connectivity, integrated access and backhaul to both small-cell and cell-free networks, and high-capacity links between satellites and ground Earth stations in the future.  NTNs improve coverage tremendously for remote regions while providing enormous capacity for urban regions and delivering services anywhere and anytime, including during disasters and emergencies.  More specifically, using High altitude platform stations (HAPSs) in tandem with TNs, space and aerial networks for various wireless applications. \textcolor{black}{For example, recently the integration of HAPS with terrestrial network \cite{Pingping_COMML_2021, Softbank}, and with satellite network \cite{Zhu_Han_JSAC_2022} have been considered for improving the network performance}. \textcolor{black}{HAPS is a quasi-stationary structure and deployed in the stratosphere at altitudes between $20 ~\mathrm{km}$ and $50 ~\mathrm{km}$ \cite{grace2011broadband}.} To reap the benefits HAPS offers, companies worldwide have started trials on HAPS-enabled wireless networks. Very recently, Stratospheric Platforms Limited (SPL), UK, and HAPSMobile, Japan, have successfully conducted trials of HAPS for ubiquitous wireless connectivity.   

\textcolor{black}{Traditionally, HAPS has been seen as a satellite competitor, especially to the low-Earth orbit (LEO) satellites. Nevertheless, neither satellites nor HAPS alone can offer networking services to future applications such as aerial cargo delivery and flying taxis. Therefore, integrated satellite-HAPS system is inevitable as it will improve the performance of satellite-specific applications, and provide deeper penetration into communications and networking-related applications \cite{Zhu_Han}}. 

This paper focuses on HAPS and its role as one of the key infrastructures of vHetNets in achieving the vision of 6G for the future. The reason for our focus on HAPS is due to its three key physical characteristics\textcolor{black}{, which make it distinct from both terrestrial BSs and satellites}: i) its strategic position relative to the Earth's surface, which allows strong line-of-sight (LoS) links \textcolor{black}{and large coverage, compared to terrestrial BSs}, lower path loss, and latency compared to satellites; ii) its size, which allows them to house large antenna arrays, computing resources and self-interference free full-duplex communications \textcolor{black}{compared to terrestrial BSs}; and iii) its quasi-stationary nature, which allows ubiquitous and seamless connectivity.

\begin{figure*}[h]
\centerline{\includegraphics[width=0.99\linewidth]{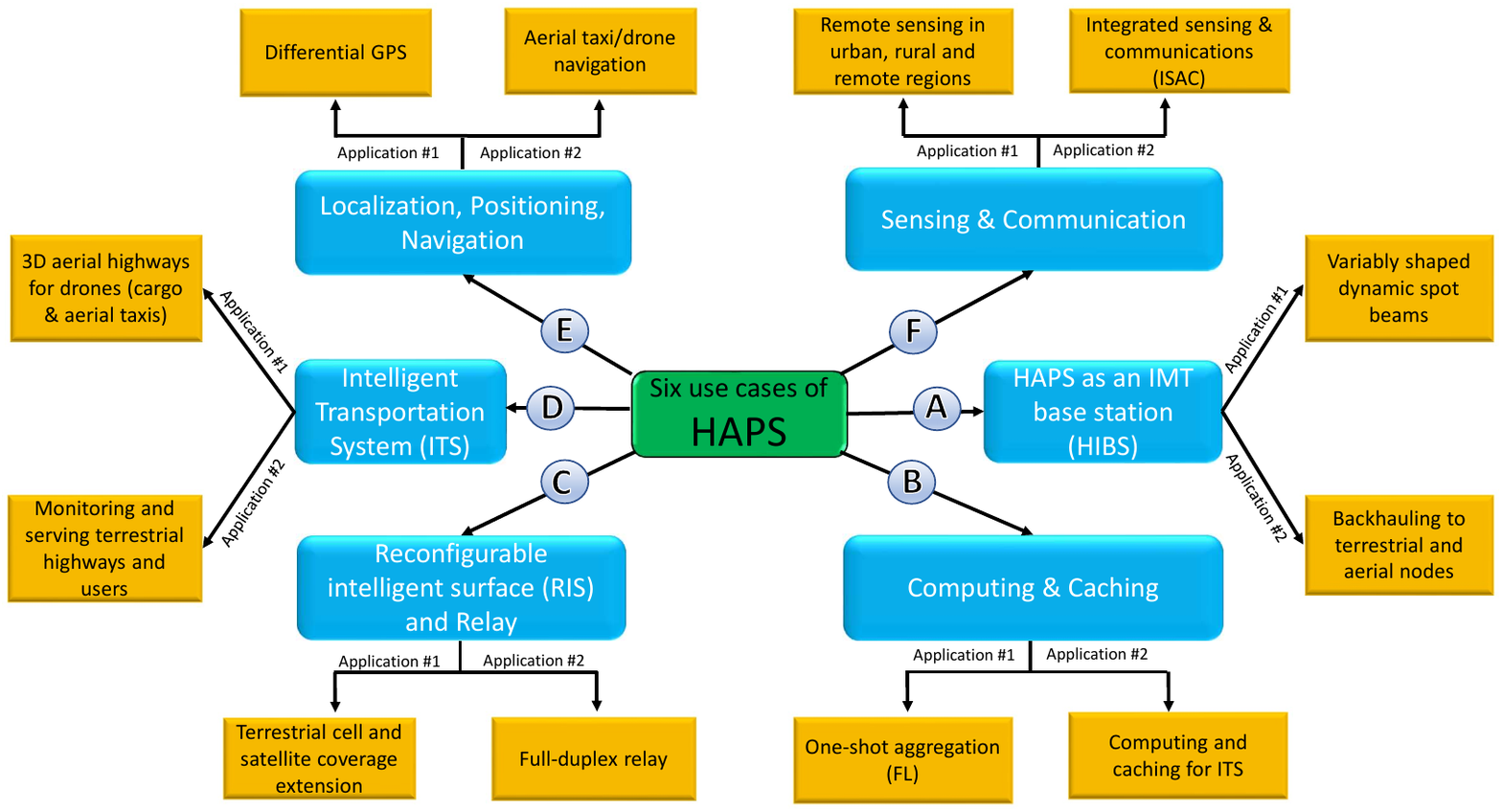}}
\caption{\textcolor{black}{\textcolor{black}{A summary of six potential use cases for HAPS and two application scenarios for each use case from a 6G perspective.}}}\label{fig:detailed_Uses_Cases}
\vspace{-0.1in}
\end{figure*}

This paper makes the following contributions:
\begin{itemize}
    \item It offers a comprehensive overview of the potential and novel use-cases where HAPS together with an existing legacy TN and space network can offer enhanced and better services. 
    \item It details several practical challenges that arise due to the physical characteristics of HAPS and its integration with existing legacy networks. Also, some possible solutions are suggested to tackle these challenges.  
    \item It presents a quantitative investigation of two use cases to verify the usefulness of HAPS-enabled vHetNets. The advantage of HAPS is demonstrated: i) on the rates of the cell-edge users, and ii) for backhauling of aerial base station.
\end{itemize}



\section{Use Cases}
In this section, we focus on HAPS as one of the key network infrastructures for providing ubiquitous connectivity and supporting dynamic capacity demand in urban regions with legacy networks. In doing so, we discuss six novel use cases relevant to 6G and beyond networks. The summary of the use cases of HAPS and its application scenarios are shown in Fig.~\ref{fig:detailed_Uses_Cases} and Fig.~\ref{fig:test}. 

\subsection{HAPS as an IMT Base Station (HIBS)}
HAPS has been envisioned in the role of a traditional international mobile telecommunications (IMT) BS, as a super macro BS (SMBS) serving different types of users in mega-cells alongside legacy macro-cell and small-cell BSs. In the international telecommunication union (ITU) terminology, a HAPS is defined as ``an IMT base station (HIBS) located on a platform that flies and stays in the stratosphere at an altitude of about 20 km.'' A HIBS can work in tandem with terrestrial BSs to support dynamic and unpredictable capacity demand in urban regions. There will be coverage holes even after the dense deployment of small-cell BSs. A HIBS can complement a TN by leveraging the HIBS's large coverage area. Furthermore, a HIBS can provide a viable solution for various unpredictable events, such as large music concerts or natural disasters, which TNs cannot handle.
\subsubsection{Variably shaped dynamic spot beams}
Employing ultra-massive multiple-input and multiple-output (UM-MIMO) technology together with higher frequency bands, such as millimeter wave (mmWave) and THz, can achieve a large capacity. At mmWave and THz bands, the RF signal suffers substantial path loss. In such frequency bands, the ultra-massive number of antennas helps to compensate the path loss by forming very narrow beams. \textcolor{black}{Further, deploying UM-MIMO on a traditional terrestrial BS is impractical, whereas, it is feasible on HAPS due to its large dimension \cite{ITU_Report_HAPS_mmWave_2000} (e.g., the Thales Alenia Space Stratobus 140 m long and 33 m in diameter \cite{stratobus2021}).} Since a HAPS can cover many users on the ground, UM-MIMO can help a HAPS form hundreds of high capacity narrow spot beams to cover one or more users on the ground.  Moreover, by creating multiple sub-antenna configurations, the size and shape of the spot beams can vary dynamically on the basis of applications and services. For example, a stadium full of people can be covered by several spot beams, and a high-speed train can be covered by one large spot beam or a narrow beam following the train to reduce the handoff.

\subsubsection{Backhauling to terrestrial and aerial nodes} \textcolor{black}{ By leveraging LoS links, a HAPS can be utilized for backhauling of both terrestrial and aerial BSs using either radio frequency (RF) or free-space optical (FSO) bands.} For terrestrial BSs, backhauling through fibre links is expensive and is impossible for dense small-cell networks. \textcolor{black}{Further}, wireless backhauling requires a lot of bandwidth, which can be achieved by using higher frequency bands (e.g., mmWave bands). However, the signals in higher frequency bands require LoS links for communication because they are highly susceptible to blockage. Thus, both wireless and fibre backhauling options are hard to implement. Alternatively, due to the existence of LoS links with terrestrial BSs, a HAPS is a suitable candidate for wireless backhauling of these BSs using higher frequency bands and FSO links.  A proof of concept for backhauling of aerial BSs via HAPS is presented in Section IV of this paper.


\begin{figure*}
\begin{adjustbox}{minipage=1\linewidth-4pt,margin=0pt 5pt,bgcolor=white,frame=1pt}
\centering
\begin{subfigure}{\textwidth}
\centerline{\includegraphics[width=0.9\linewidth]{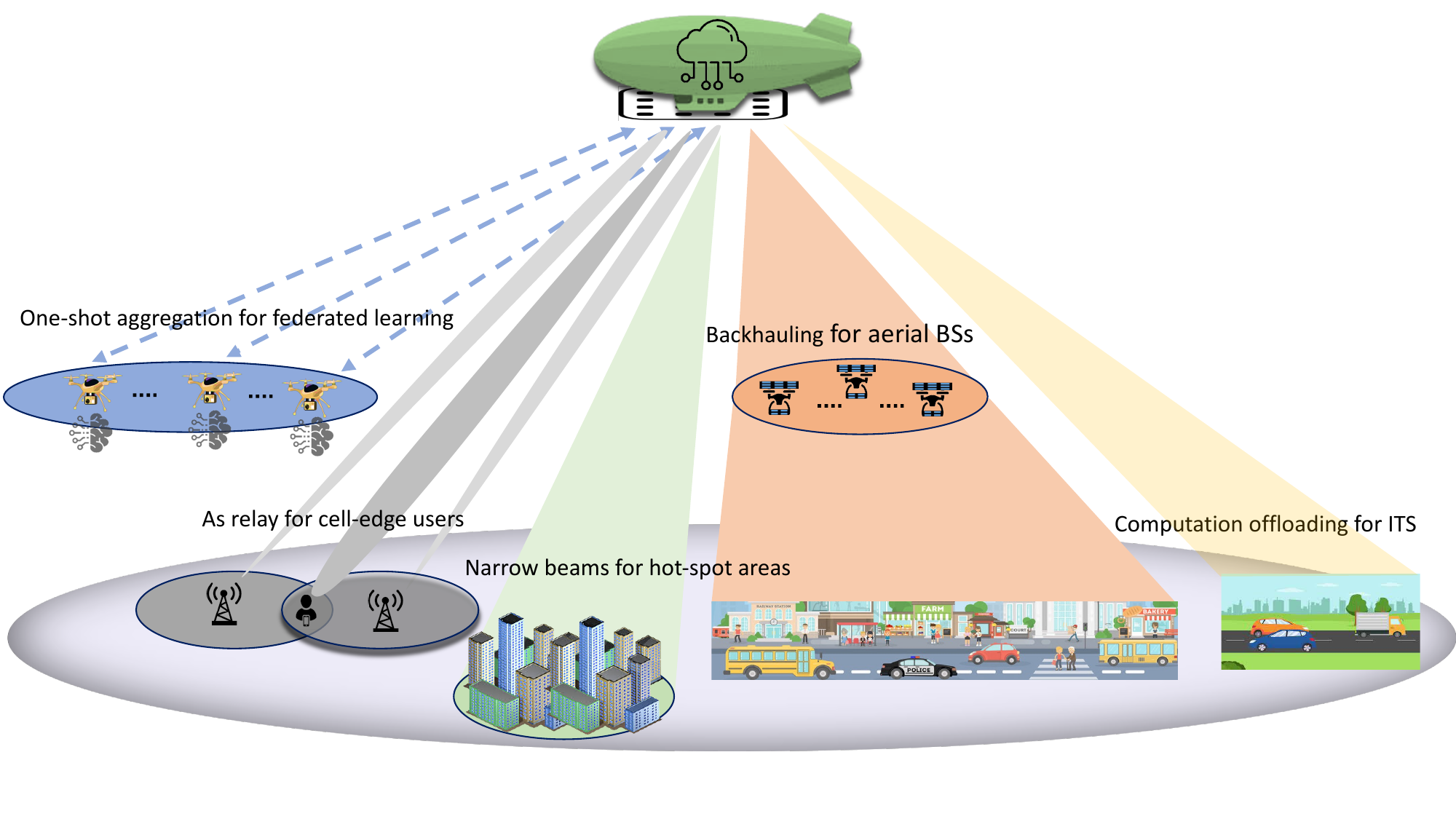}}
\caption{}
\label{fig:overview_use_cases}
\end{subfigure}
\begin{subfigure}{.5\textwidth}
  \centering
  \includegraphics[width=\linewidth]{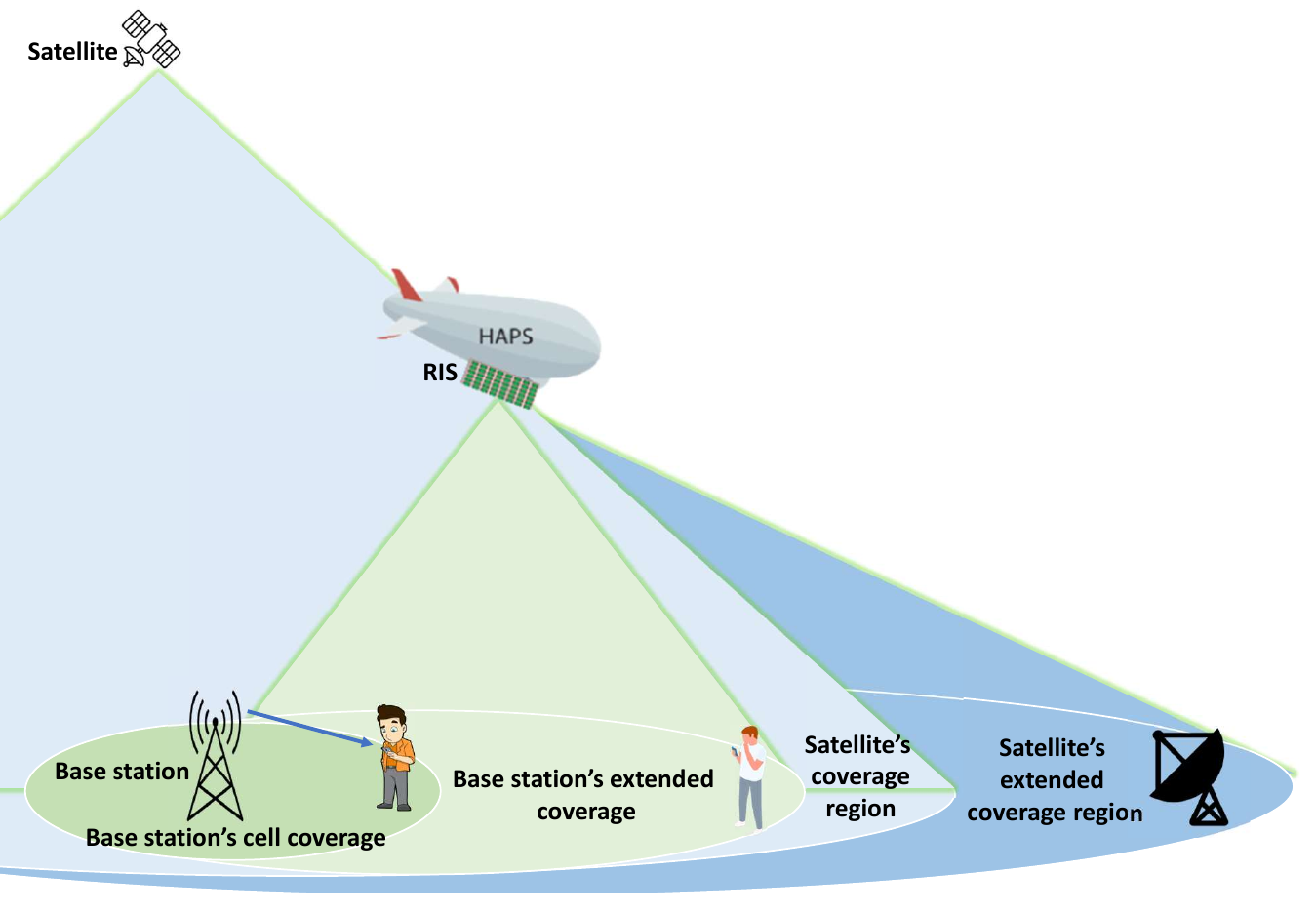}
  \caption{}
  \label{fig:cell_extension}
\end{subfigure}%
\begin{subfigure}{.5\textwidth}
  \centering
  \includegraphics[width=\linewidth]{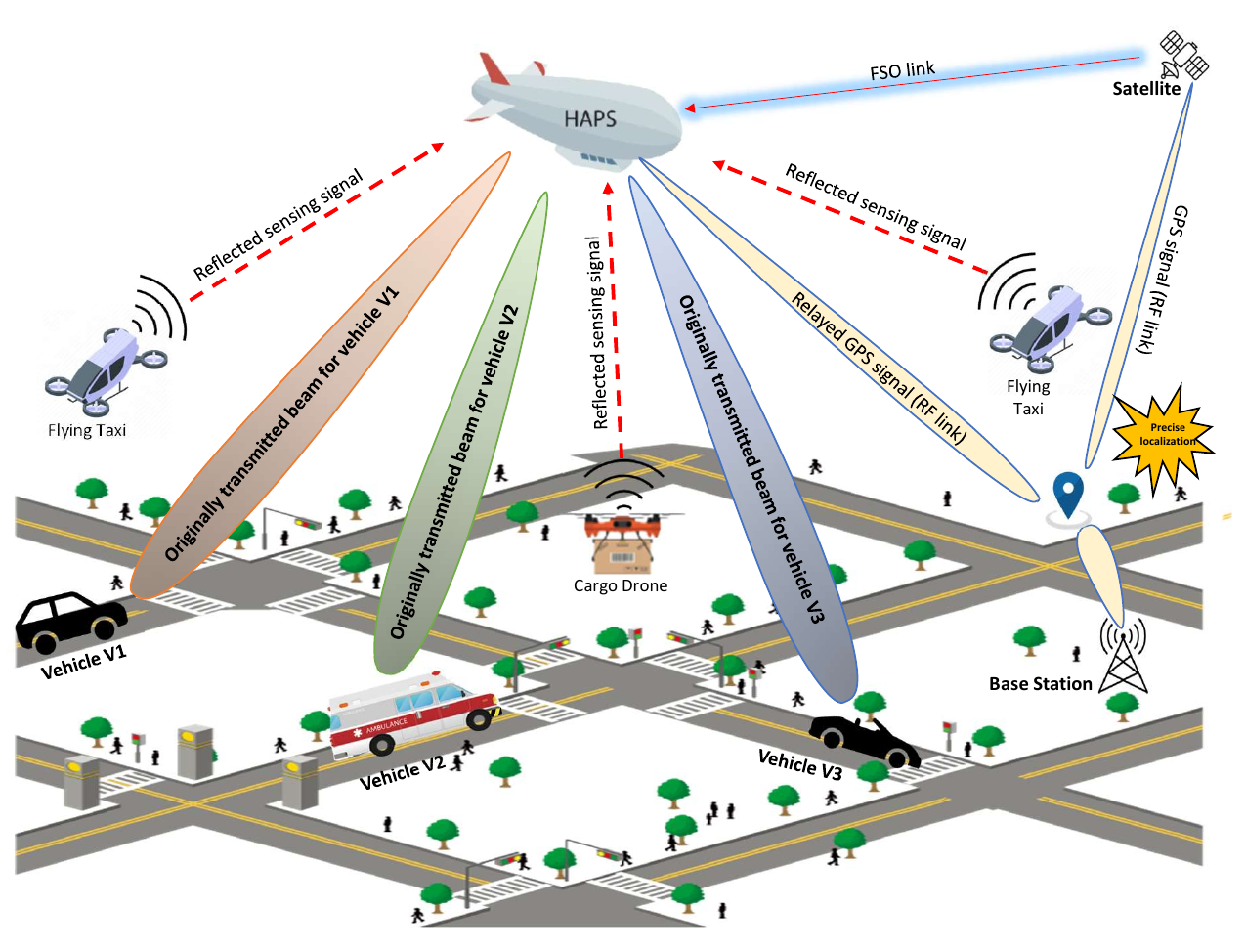}
  \caption{}
  \label{fig:ISAC_GPS}
\end{subfigure}
\caption{\textcolor{black}{(a) Application scenarios of HAPS for one-shot aggregation at federated learning, as a relay for cell-edge users, to provide narrow beams for hot-spot areas, for backhauling of aerial BSs, and for computation offloading at ITS. (b) Coverage extension via RIS-mounted HAPS. (c) Integrated sensing and communications (ISAC), and localization via HAPS.}}
\label{fig:test}
\end{adjustbox}
\end{figure*}
\subsection{Computing and Caching}
\subsubsection{Computing and caching for ITS}
Due to physical limitations and energy-supply constraints, the computing capabilities of edge equipment are usually limited. HAPS computing can be considered as a promising extension of edge computing. In this regard, it is suitable to coordinate terrestrial resources and store the common data associated with ITS-based applications. Three computing layers (i.e., vehicular, terrestrial network edge, and HAPS) can be integrated to build a computation framework for ITS, where the HAPS data library stores the data needed for the applications. In addition,  by utilizing a caching technique at the network edges, we can store some of the common data from the HAPS so that large transmission delays can be reduced. 
\subsubsection{One-shot aggregation}
Federated learning (FL) is a secure data-driven approach that deals with learning a global machine learning model by aggregating the model parameters learned on several small databases located apart. A designated centralized server is assigned to aggregate the model parameters and then broadcast the aggregated parameters to the local servers. The main idea of FL is to preserve privacy among the databases. The \textcolor{black}{convergence and} accuracy of the final model improves with the number of participating servers. But when FL takes place through terrestrial wireless networks, it can suffer from a limited number of participants in the vicinity of the centralized server. This limitation can be overcome by performing multi-tier FL at the cost of a longer convergence time due to multi-hop communications. This bottleneck can also be overcome by moving the centralized server to the HAPS. Owing to the extensive ground coverage of the HAPS, moving the centralized server to the HAPS can improve the number of participants and can communicate with each of them in a single hop.

\subsection{HAPS with Reconfigurable Intelligent Surface (RIS)}
RISs have attracted much attention in recent years. An RIS is built up of many reflecting units capable of changing the phase of the impinging RF signals, and thus it can help wireless communication by providing favorable channel conditions. However, deploying RISs in urban areas might not meet the dynamic demands. Hence, integrating RIS with an aerial surface is also getting a lot of attention \cite{Safwan}. Due to the large surface area of a HAPS, RIS units can be mounted on a HAPS to reflect and direct received signals from ground sources towards desired destinations. 
\subsubsection{Terrestrial-cell and satellite coverage extension} \textcolor{black}{In the future, urban areas will experience tremendous volume of data traffic and users}. Consequently, terrestrial BSs will become overloaded and unable to support all users. In this situation, under-loaded BSs located further away \textcolor{black}{in sub-urban or remote regions can offer service to unsupported users. The far distance connection can be achieved by mounting RIS on HAPS. The phases of reflective elements of RIS are} tuned to direct the signal coming from the BSs towards the unsupported users. \textcolor{black}{Thus, through signal reflection from RIS on HAPS can extend the coverage of terrestrial BSs}. Satellites can also play a role in such a situation, as they have a larger footprint than HAPS due to their higher altitudes. However, satellites also suffer from coverage gaps. To this end, a HAPS-RIS can be useful in widening the coverage region of satellites \textcolor{black}{by redirecting (via phase tuning of RIS elements) the satellite signals towards its coverage gap regions.}   Fig.~\ref{fig:cell_extension} shows how an RIS-mounted HAPS can extend the coverage of both terrestrial cells and satellites.

\subsubsection{Full-duplex (FD) relay} In urban areas, cell-edge users suffer from severe capacity crunches due to shadowing and blockages from high-rise buildings and other city infrastructure. This capacity crunch can be solved by leveraging the LoS links between a HAPS and ground nodes. In addition, leveraging a HAPS's large size, full-duplex communication is possible by installing transmitter and receiver antennas far apart to lower the effect of self-interference to a very low value. In this case, if the terrestrial BSs relay the signals of cell-edge users through a HAPS, the capacity of the cell-edge users can be enhanced. A proof of concept for this case is presented in Section IV. 


\subsection{Intelligent Transportation System (ITS)}
\subsubsection{Monitoring and serving terrestrial highways and users}
Autonomous decision-making systems embedded in connected and autonomous vehicles (CAVs)  create substantial amount of deep learning-based computations. Assisting CAVs to perform AI tasks at edge and cloud servers has attracted much attention recently. A HAPS with substantial computing capability can help CAVs to perform their computing tasks. Other than rich computing resources, HAPS has many advantages compared with TNs in managing the mobility of vehicles. Due to a very big coverage area, a HAPS can reduce the number of handovers significantly. 
\subsubsection{Supporting 3D aerial highways for cargo drones} Terrestrial BSs need to tilt their antennas upward to serve aerial users such as cargo drones or airplanes. A HAPS, for its part, can view all terrestrial and aerial users without having to change its antenna configuration, and thus it can provide 3D coverage for aerial and terrestrial highways. In addition, due to blockages and shadowing, the link between a terrestrial BS and a mobile aerial user is very unreliable. However, there is less probability of blockage and shadowing for the links between ground and aerial users and HAPS.
 
\subsection{Localization, Navigation, Positioning} 
\subsubsection{HAPS-assisted differential global navigation satellite system (GNSS)} Due to its lower altitude, a HAPS can help legacy localization, navigation, and positioning systems for more accurate measurements. The current GNSS estimates the position of a ground node based on four MEO satellites placed at an altitude of 20,000 km. However, a HAPS is placed at an altitude of 20 km, which can provide additional information to GNSS for accurate positioning. For example, via differential GNSS, as depicted in Fig~\ref{fig:ISAC_GPS}, satellite, HAPS, and terrestrial BS work together to deliver accurate positioning information. 
\subsubsection{3D navigation for aerial taxis and drones} It is envisaged that aerial taxis and cargo drones will be part of our daily life in the future \cite{pan2021IEEEaccess}. However, unlike road taxis and traditional cargo deliveries, aerial taxis and cargo drones, for optimal performance, will require continuous connectivity for navigation and positioning. Therefore, HAPS with its LoS links and enormous footprint will be able to offer continuous connectivity, 3D positioning, and navigation.

\subsection{Sensing and Communications} 
\subsubsection{Remote sensing} As HAPS can establish LoS links with both aerial and ground nodes over a broad coverage area, it is desirable for sensing alone applications, such as monitoring with high-resolution cameras, remote sensing with \textcolor{black}{light detection and ranging (LIDAR)}, or internet-of-thing (IoT) devices. Note that satellites are very far for these sensing applications and their provided service can not be accurate. Also, some sensors do not have the energy required to send their signals to satellites. Furthermore, sensing with terrestrial nodes is beset by a lack of LoS links with nodes and a small coverage area. HAPS is a promising candidate for ITS as it can sense vehicular signals along remote highways as well as in urban areas.
\subsubsection{Integrated sensing and communications (ISAC)} Future generation of wireless networks will encounter severe spectrum congestion and scarcity. Therefore, two different types of systems (i.e., sensing and communication systems) are required to coexist. This concept is also called ISAC. In this regard, HAPS will play a vital role in supporting such coexistence. For example, consider an integrated passive localization and communication system, as depicted in Fig.~\ref{fig:ISAC_GPS}. The HAPS sends information signals to three mobile ground users, and at the same time, these information signals are also reflected by three drones in the area. The reflected signals received by the HAPS can be used to localize the drones. Thus, sensing and communications can take place simultaneously, which improves the spectrum utilization.

\begin{table*}[h!]
	\caption{The summary of challenges of integrating HAPS into 6G and their solutions.}
	\begin{center}
	{\rowcolors{2}{white}{black!10}
		\begin{tabular}{|m{9cm}| m{8cm}| }
			\hline 
		\textbf{Challenge} & 	\textbf{Suggested Solution} \\ 
			\hline \hline

   \textbf{Designing UM-MIMO architecture:} \begin{itemize}
		    \item Substantial energy consumption of RF chains \textcolor{black}{in mmWave and THz bands}. 
		    \item Pilot contamination due to substantial number of beams and limited reference signals.
		    \item Inter-beam interference due to relatively large footprint of each beam.
		    \item Inter-user and inter-element correlation.
		\end{itemize} & \begin{itemize}
		    \item Applying analog or hybrid beamforming techniques to reduce the number of RF chains.
		    \item Performing RRM techniques to control interference.
		    \item User grouping to manage inter-user correlation.
		    \item Increasing antenna element spacing to control inter-element correlation.
		\end{itemize}\\ 

 \textbf{TN-HAPS-LEO integration:} \begin{itemize}
		    \item 	Need for efficient RRM and multi-band connectivity through the dynamic use of FSO, mmWave, and THz technologies.  
		    \item Heterogeneity in energy availability at terrestrial and non-terrestrial networks.
		    \item Need for a seamless integration for continuous connectivity. 
		\end{itemize} & \begin{itemize}
		    \item Developing rapidly converging and decentralized algorithms that can be implemented on GPUs.
		    \item Machine learning can be utilized for efficient optimization of all tiers.
		\end{itemize}\\ 
  
		\textbf{Backhauling of HAPS:}
		\begin{itemize}
		    \item Need for a high capacity backhaul link to connect HAPS to the core network.
		    \item Need for designing transceivers at both the HAPS and ground station that are robust enough to pointing errors and changes in weather conditions.
		\end{itemize}& 	\begin{itemize}
		    \item Utilizing high bandwidth in THz band or FSO link to forward the signals of HAPS to the ground BS. 
		\end{itemize} \\ 
			\textbf{Jittering in HAPS placement:}  \begin{itemize}
		    \item Caused by wind in the stratosphere.
		    \item  Leads to a lower coherence time for the HAPS to user channels.
		    \item Need for deploying advanced channel estimation schemes to track these channels. 
		   
		    \item Leads to laser pointing error for inter-HAPS communication with FSO.
		    \item Leads to variation in the HAPS's footprint, and hence beam misalignment and cell-edge user disconnectivity.
		\end{itemize} & \begin{itemize}
		   
		    \item Using a stale CSI technique to solve channel estimation problem, where the stale and current CSI are combined softly. 
		    \item Deploying efficient jitter-aware beam tracking and handover algorithms to address moving cell footprint problem. 
		\end{itemize}\\ 
			\textbf{Limited onboard energy:} \begin{itemize}
		    \item Due to serving a large number of terrestrial and aerial vehicles.
		    \item Not continuous power supply, like terrestrial power grid.  
		\end{itemize} &  \begin{itemize}
		     \item Renewable energy sources, such the sun and wind. 
		     \item Mid-air refuelling, replacing batteries regularly, and powering via tethering.
		     \item Developing energy-efficient communications protocols from the signal processing viewpoint\textcolor{black}{, and leveraging strong LoS links for low energy communication links.}
		\end{itemize}\\ 
			\textbf{HAPS location estimation:} \begin{itemize}
		    
		    \item Due to the jittering of a HAPS, its location changes a lot.
		    \item In HAPS-assisted differential GNSS for positioning nodes, incorrect HAPS location information adds randomness to the GNSS information.
		   
		\end{itemize} & \begin{itemize}
		    \item RL-based algorithms that can regularly predict HAPS's exact location by feeding atmospheric data such as wind velocity. 
		    \item The HAPS can estimate its exact location by comparing its current position with a reference beacon on the ground.
		\end{itemize}\\ 

		\textbf{HAPS with RIS:} \begin{itemize}
		    \item 	RIS units together need a large amount of energy for operation.
		     \item How to use the RIS to serve multiple areas simultaneously.
		    \item RIS-mounted HAPS has no fibre option available to estimate channels like the one available for terrestrial RIS. 
		\end{itemize} & \begin{itemize}
		    \item Designing energy-efficient approaches for he operation of RIS elements.
		    \item Developing efficient channel estimation schemes is the most crucial signal processing step.
		\end{itemize}\\ 

  		\textbf{Security:} \begin{itemize}
		    \item 	\textcolor{black}{ Vulnerable to attacks due to wireless nature of HAPS communication. For example, malicious participants can join the FL intending to reduce the global model's accuracy.}
		    \item \textcolor{black}{ Having only one centralized server at the HAPS for many participants poses a security problem. For example, the whole FL system collapses if the HAPS becomes an anomaly. }
		    
		\end{itemize} & \begin{itemize}
		   \item \textcolor{black}{ Using ML models to detect abnormal behaviors in the network can be a solution to secure critical
infrastructures.}
   \item \textcolor{black}{ To ensure the secure information transmission,
a designed noise can be added to the local model parameters in FL to achieve
differential privacy.}
		\end{itemize}\\ 	
		
		\hline
		
		\end{tabular}}\label{tab:challanges_and_solutions}
	\end{center}
\end{table*}

\section{Challenges and Solutions}
To reap the benefits of employing a HAPS for the aforementioned use cases, we need to address some practical challenges. In this section, we highlight some important challenges associated with integrating HAPS into future wireless networks, and we discuss potential solutions to address them. The summary of challenges and their solutions is presented in Table~\ref{tab:challanges_and_solutions}. 
\subsection{Designing UM-MIMO Architecture}
There are some practical challenges for HAPS antenna architecture. The first issue is due to the substantial energy consumption of RF chains utilized for an ultra massive number of antenna elements. In order to solve this problem, analog or hybrid beamforming techniques can be utilized to reduce the number of RF chains and their energy consumption.   Another issue is that of pilot contamination in HAPS, which is due to a multiplicity of beams and limited pilot reference signals. Thus, applying the same pilot reference for more than one user leads to this issue.
The third problem is inter-beam interference due to the relatively large footprint of each beam which can be addressed by radio resource management techniques. Finally, inter-user and inter-element correlation can impact the performance of a HAPS system. User grouping can be utilized to manage inter-user correlation and the spacing between antenna elements can be increased to control inter-element correlation.

\subsection{HAPS Integration with Satellite and Terrestrial Networks}
The marriage between HAPS, TN, and satellite networks is inevitable. Together, they can provide global high-capacity coverage using a single system. There are many challenges related to efficient radio resource management (RMM) and multi-band connectivity through the dynamic use of FSO, mmWave, and THz technologies \cite{tashiro2022ieeeaccess}. What is required are decentralized algorithms that converge rapidly and that can be implemented on general-purpose units (GPUs). Further, there is heterogeneity in energy availability at terrestrial and non-terrestrial networks. Because of this, seamless integration for continuous connectivity and more prolonged operations are always a bottleneck. Machine learning can be utilized for the efficient optimization of all tiers.

\subsection{Backhauling of HAPS} When a HAPS operates in the capacity of a HIBS, a high-capacity backhaul link is required for its seamless connectivity to the core network. This high-capacity backhauling can be achieved by connecting the HAPS to a ground station using RF or FSO links.  High bandwidth in the THz frequency band or an FSO link are excellent options for backhauling. However, they need the development of novel transmitter and receiver schemes at both the HAPS and the ground station that are robust to pointing errors and changing of the weather conditions. It should be noted that the HAPS can work as both a transparent node or regenerative node to forward the received signals from users to the core network. In contrast to satellites, which need to connect to many ground stations as they orbit, each HAPS requires only one ground station for backhauling.

\subsection{Jittering in HAPS Placement} Besides the movements of terrestrial users, HAPS experiences jitter due to wind in the stratosphere. The compounded effect of user movement and jitter leads to a lower coherence time for the channel between the HAPS  and the user. Thus, developing advanced channel estimation schemes to track these channels is required. Using a stale \textcolor{black}{channel state information (CSI)} technique can solve this problem, where the stale and current CSI are combined softly. The jittering of a HAPS can also impact inter-HAPS communication, especially when FSO is utilized to connect HAPSs and the laser pointing errors lead to information loss. Furthermore, jitter also leads to variation in a HAPS's footprint. As a result, severe consequences such as beam misalignment and cell-edge user disconnectivity may occur. Thus, efficient jitter-aware beam tracking and handover algorithms are needed.

\subsection{Limited Onboard Energy} Since a HAPS is in the stratosphere, there is no continuous power supply like there is for terrestrial BSs. Renewable energy sources, such as the sun and wind, are the obvious solutions. Besides, mid-air refuelling, replacing batteries regularly, and powering via tethering are some alternative solutions to keep a HAPS operational without discontinuity. Moreover, developing energy-aware and energy-efficient communications protocols and algorithms from a signal processing viewpoint becomes vital.   

\textcolor{black}
{Note that HAPS can provide additional capacity that can be
used in a flexible way to facilitate terrestrial BSs’
sleep modes and, ultimately, reduce energy consumption and make the network more sustainable \cite{Sustainable_HAPS}.
Indeed, offloading
traffic to HAPS allows reduction of
the grid energy demand of terrestrial nodes while
still maintaining adequate quality of service.
}

\subsection{HAPS Location Estimation} 
In theory, a HAPS's location is known; however, its location keeps changing due to jitter. \textcolor{black}{Knowing the exact location of HAPS is essential for the optimal performance of the previously discussed use cases. For example, knowing the exact location of HAPS in HAPS-assisted differential GNSS improves the overall localization accuracy compared to that of standalone GNSS. However,} incorrect HAPS location information adds randomness to the GNSS information and leads to an even worse outcome. To address the HAPS localization problem, we can add other positioning systems to estimate the exact location of the HAPS. For example, reinforcement learning (RL)-based algorithms can regularly predict a HAPS's exact location by feeding atmospheric data such as wind velocity. In another method, the HAPS can estimate its exact location by comparing its current position with a reference beacon on the ground. Furthermore, a better estimate of the HAPS's location can be obtained by combining the two approaches. 

\subsection{HAPS with RIS}
A few challenges must be addressed when RIS units are mounted on a HAPS. Firstly, due to the large surface area of a HAPS, an RIS with a large number of reflective units can be mounted on its hull. However, each unit requires energy for operation, and all RIS units together need a large amount of energy. To this end, the design of energy-efficient approaches is necessary. Secondly, how to use the RIS to serve multiple areas simultaneously is an important problem that needs an efficient design solution. In addition, an RIS-mounted HAPS has no fibre option available to estimate channels like the one available for terrestrial RISs. Thus, developing efficient channel estimation schemes is the most crucial signal processing step that needs to be addressed to reap the benefits of HAPS with RIS. 

\subsection{Security}
Due to the wireless nature of HAPS communication, it is vulnerable to security attacks. For example, due to the large number of local servers in the FL process, malicious participants can join the FL intending to reduce the global model's accuracy. In addition, having only one centralized server at the HAPS for many participants poses a security problem. Again, in the FL example, the whole FL system collapses if the HAPS becomes an anomaly. 

\textcolor{black}{As an emerging decentralized learning framework, FL is effective to overcome the challenges of privacy
concern by enabling
distributed clients to train a shared anomaly detection model locally without data centralization.
However, since the model parameters
of all clients are required to exchange with HAPS through
wireless links periodically in the model training process,
the sensitive information is more likely to be inferred. Using ML models to detect abnormal behaviors in the network can be a solution to enhance security.
Furthermore, to ensure the secure information transmission,
a designed noise can be added to the local model parameters.}
\begin{table}[b]
	\caption{Simulation parameters.}
	\vspace{-11pt}
	\begin{center}
	{\rowcolors{2}{white}{black!10}
		\begin{tabular}{|c|c|} \hline \label{sim_param}
		\textbf{Parameter} & 	\textbf{Value} \\ 
			\hline \hline
		Carrier frequency of the first hop & 2 GHz \\ 
			Carrier frequency of the second hop & 120 GHz \\ 
			Communication bandwidth & 1 MHz \\ 
			Noise power spectral density & -174 dBm/Hz\\ 
			 Number of users & 16 \\ 
			Number of UxNBs & 16 \\ 
			Number of receive antenna elements at each UxNB & 4 \\ 
				Transmit power at each user & 23 dBm \\ 
			Transmit power at each UxNB  & 25 dBm \\ 
			Absorption coefficient of the sub-THz medium & 0.5 dB/km\\
			Fixed flight height of all UxNBs & 120 meters \\
			Altitude of HAPS & 20 km\\
			\hline
		\end{tabular}}
	\end{center}
	\vspace{-19pt}
\end{table}

\textcolor{black}{Undoubtedly, HAPS can perform multiple functions simultaneously, such as IMT BS (HIBS), HAPS-RIS, and federated learning agent. For a successful multi-functional operation, multi-agent ML can play a vital role where learning for each function can be performed on orthogonal computing and communication resources of HAPS. }

\section{Performance Studies}
\textcolor{black}{This section qualitatively examines the integration of HAPS with terrestrial and aerial networks. Particularly, we consider two use cases: (i) HAPS as a relay, and (ii) HAPS as a HIBS, to study the performance achieved by the HAPS-integrated networks.}
\subsection{Capacity Enhancement of Cell-Edge Users with HAPS}
\textcolor{black}{In conventional terrestrial downlink networks cell-edge users suffer from capacity loss due to substantial shadowing and intercell interference. To overcome this challenge, in} the first Case Study, we consider cell-edge users' capacity enhancement in a two-cell scenario supported by full-duplex HAPS. The two BSs in each terrestrial cell and HAPS employ 64 and 100 antenna elements, respectively, in the uniform plannar array (UPA) configuration and operate in the 28 GHz band. Two single antenna users in each cell are considered, with one user being close to the BS and the other being far, at the cell edge. The BSs and HAPS employ fully-connected hybrid precoder architecture with two RF chains each.
In contrast, the HAPS employs a fully analog combiner for receiving the signals.  \textcolor{black}{The signal setup is as follows: the each BS beamforms the composite signal of the near and cell-edge users using the same time-frequency resource towards the near user and FD-HAPS. The FD-HAPS, after receiving the signals from two BSs, decodes the signals of the two edge users and then beamforms towards the cell-edge users.} The sum rate of the considered system is maximized by jointly designing the hybrid beamformers for both the BSs and the HAPS. The sum rate is averaged over the 200 i.i.d., generated random topologies and 500 i.i.d., channel realizations for each topology. 
The terrestrial cells, each with a radius of 300 m, are assumed to be set up in a typical urban area.

Fig. \ref{fig:cell-edge} presents a comparison \textcolor{black}{of the  cumulative density functions (CFDs)} of the system's total sum rate obtained by joint design with HAPS, without HAPS, and selfish design (no BS coordination). It can be observed that the fifth percentile that corresponds to the edge users is higher than for the cases where HAPS is absent.   

\subsection{Backhauling for Aerial BSs with HAPS}
In the second Case Study, we use a HAPS as a central processing unit (CPU) for backhauling aerial BSs (UxNBs) that are working in cell-free mode. At the first time-slot, users send their messages to UxNBs at the sub-6 GHz band. The UxNBs then apply match-filtering and power allocation. At the second time-slot, at each UxNB, orthogonal resource blocks are allocated for each user at the sub-THz band, and the signals are sent to the HAPS after analog beamforming. In the HAPS receiver, after analog beamforming, the message of each user is decoded. We found the optimum values for the allocated power for users in each UxNB that maximizes the minimum SINR of users. For the simulations, we assume that all users are uniformly distributed over a square urban area with a length of $1,000$ meters. Also, we assume that the HAPS is deployed in the middle of this square area. One can see the simulation parameters in Table \ref{sim_param}. 



\begin{figure}[h]
\begin{adjustbox}{minipage=1\linewidth-4pt,margin=0pt 5pt,bgcolor=white,frame=1pt}
\centering
\begin{subfigure}{\textwidth}
  \centering
  \includegraphics[width=0.85\linewidth]{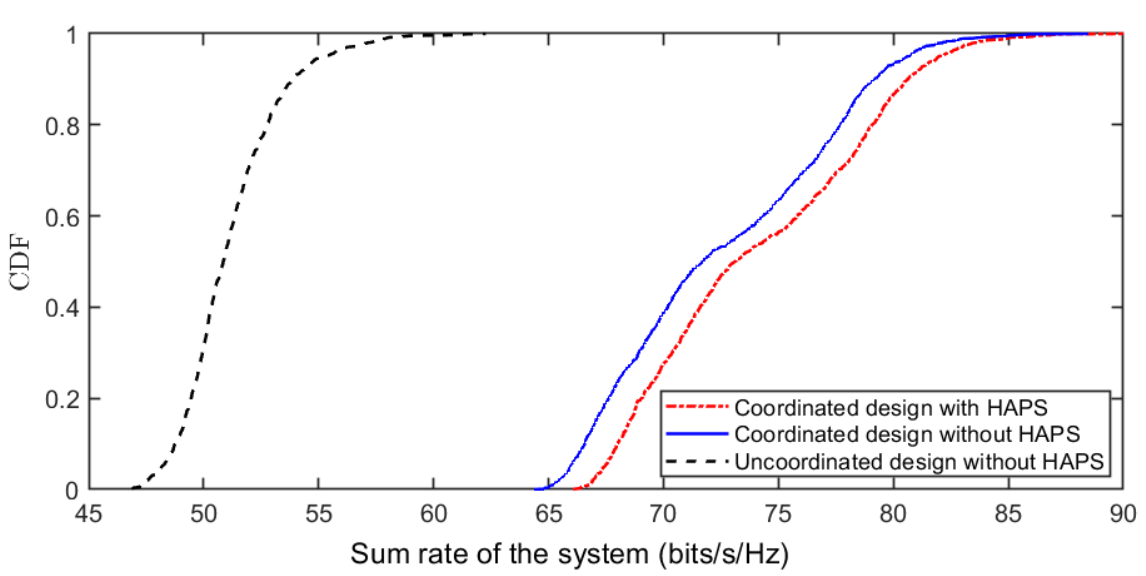}
  \caption{}
  \label{fig:cell-edge}
\end{subfigure}%
\newline
\begin{subfigure}{\textwidth}
  \centering
  \includegraphics[width=1\linewidth]{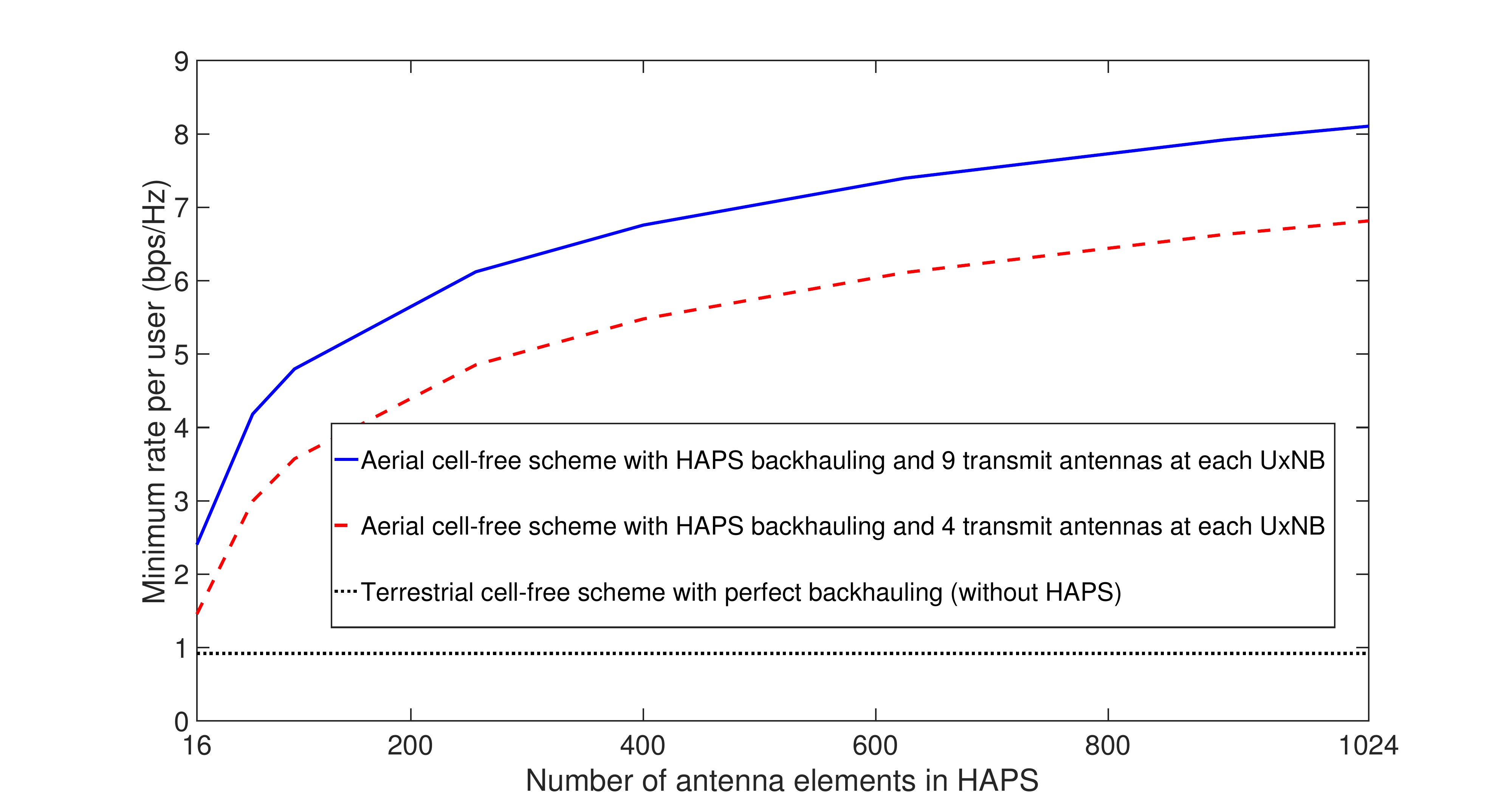}
  \caption{}
  \label{R_vs_S_P25_M16_K16}
\end{subfigure}
\caption{(a) Cumulative distribution function (CDF) versus sum rate of the system for Case Study 1. (b) The achievable minimum rate per user versus the number of HAPS antenna elements for Case Study 2.}
\label{fig:sim_res}
\end{adjustbox}
\end{figure}

In Fig. \ref{R_vs_S_P25_M16_K16}, it can be observed that when the number of HAPS antenna elements is high, aerial schemes have a significant performance gain over the terrestrial cell-free scheme\footnote{In terrestrial cell-free baseline scheme, each terrestrial user is served by multiple terrestrial access points, and a perfect backhaul with fiber links is assumed to connect the access points and CPU.}. \textcolor{black}{It should be emphasized that in the terrestrial cell-free scheme, the links between users and access points (APs) are NLoS, and hence the path loss is high. However, in the proposed aerial cell-free scheme, since there is a strong LoS link between the users and UxNBs, the path loss is lower.} This figure shows that utilizing a HAPS as a CPU is useful when the enormous path loss between the UxNBs and the HAPS in the sub-THz band is compensated for by a high number of antenna elements at the HAPS.

\section{Conclusion}
This article has focused on HAPS as a vital infrastructure node of VHetNets envisioned for 6G and beyond wireless networks.
 Relevant use cases pertaining to 6G and beyond networks were discussed, where HAPS could work in tandem with legacy networks. Some important practical open challenges associated with HAPS and its integration with legacy networks were discussed. Furthermore, some potential solutions were suggested to address the challenges. Through numerical simulations, two  use cases were analyzed to validate the superiority of integrating HAPS into terrestrial network.

\vspace{-.45in}
\begin{IEEEbiographynophoto}{Omid Abbasi}
\textbf{[SM]} (omidabbasi@sce.carleton.ca) is currently with the Department of SCE at Carleton University. He is a Senior Member of IEEE.
\end{IEEEbiographynophoto}
\vskip -2\baselineskip plus -1fil
\begin{IEEEbiographynophoto}{Animesh Yadav}
\textbf{[SM]} (animesh.yadav@mnsu.edu) is an Assistant Professor in the Department of ECET at Minnesota State University, Mankato, MN, USA. He is a Senior Member of IEEE.
\end{IEEEbiographynophoto}
\vskip -2\baselineskip plus -1fil
\begin{IEEEbiographynophoto}{Halim Yanikomeroglu}
\textbf{[F]} (halim@sce.carleton.ca) is a Full Professor in the
Department of SCE at Carleton University.
His research focus is on aerial and satellite networks for the
6G era. 
He is a Fellow of IEEE, and a Distinguished Speaker for both IEEE
Communications Society and IEEE Vehicular Technology Society.
\end{IEEEbiographynophoto}
\vskip -2\baselineskip plus -1fil
\begin{IEEEbiographynophoto}{Ngoc Dung Dao}
(ngoc.dao@huawei.com) is a principle engineer of
Huawei Technologies Canada. His research interest covers several aspects
of 5G and beyond 5G mobile networks, including architecture design, data
analytics, and vertical applications. 
\end{IEEEbiographynophoto}
\vskip -2\baselineskip plus -1fil
\begin{IEEEbiographynophoto}{Gamini Senarath}
\textbf{[SM]} (gamini.senarath@huawei.com) is
with 5G Research Lab, Huawei Technologies, Canada, and also the Project Manager for Huawei external collaboration projects. His current research interests include
5G wireless systems.
\end{IEEEbiographynophoto}
\vskip -2\baselineskip plus -1fil
\begin{IEEEbiographynophoto}{Peiying Zhu}
\textbf{[F]} (peiying.zhu@huawei.com) is a Huawei Fellow, IEEE Fellow and Fellow of Canadian Academy of Engineering.
She is
currently leading 5G wireless system research in Huawei. 
The focus of her research is advanced wireless access technologies. 
\end{IEEEbiographynophoto}

\end{document}